\documentclass[aps,prb,reprint,showpacs,groupedaddress]{revtex4-1}

\usepackage{graphicx}
\usepackage{dcolumn}
\usepackage{bm}
\usepackage{amsmath,amssymb}

\def\U#1{{%
\def\O{\mbox{O}}
\def\u{\mbox{u}}
\mathcode`\u=\mu
\mathcode`\O=\Omega
\mathrm{#1}}}

\def\dd{{\mathrm{d}}}

\begin{document}


\title{Observation of modulation instability in a nonlinear
magnetoinductive waveguide}

\author{Yasuhiro Tamayama}
\email{tamayama@vos.nagaokaut.ac.jp}
\altaffiliation[present address: ]{Department of Electrical Engineering,
Nagaoka University of Technology, 1603-1 Kamitomioka, Nagaoka, Niigata 
940-2188, Japan}
\author{Toshihiro Nakanishi}
\author{Masao Kitano}
 \affiliation{Department of Electronic Science and
 Engineering, Kyoto University, Kyoto 615-8510, Japan}

\date{\today}

\begin{abstract}
We report numerical and experimental investigations into
modulation instability in a nonlinear magnetoinductive waveguide. By
numerical simulation we find that modulation instability occurs in an electrical
circuit model of a magnetoinductive waveguide with
third-order nonlinearity. We fabricate the nonlinear
magnetoinductive waveguide for microwaves using varactor-loaded
split-ring resonators and observe the generation of modulation
instability in the waveguide. 
The condition for generating
modulation instability in the experiment roughly agrees with that in
the numerical analysis.
\end{abstract}

\maketitle

\section{Introduction}

Considerable effort has been devoted to controlling electromagnetic waves
using metamaterials that consist of arrays of subwavelength structures. 
The macroscopic characteristics of metamaterials depend on the structure of
the constitutive elements as well as the 
characteristics of the materials. It is therefore 
possible to design artificial media with extraordinary properties by 
devising the structure and material of the constitutive elements. 
Various kinds of unusual 
media and phenomena have been realized using
metamaterials.\cite{smith04_sci,alitalo09,wang_b09_jopt,lukyanchuk10,soukoulis11,zheludev12} 
 
Efficient generation of nonlinear waves is 
a promising application of metamaterials.\cite{pendry99} 
When an electromagnetic wave is incident on a metamaterial composed of
resonant elements such as split-ring resonators, the electromagnetic
energy is compressed into a small volume at certain locations of 
the element at the resonant
frequency. If nonlinear elements are placed at these locations, 
nonlinear phenomena occur efficiently. 
Several nonlinear phenomena such as harmonic
generations,\cite{klein06,shadrivov08_apl,kim08,kanazawa11,nakanishi12,reinhold12} 
bistability,\cite{wang_b08} and tunable
media\cite{shadrivov08_apl,wang_b08,powell09} have been experimentally
investigated. 

In this paper, we focus on modulation instability. 
Modulation instabilities are observed in various nonlinear systems.
In the field of optics, modulation instabilities in optical fibers 
have been actively investigated. 
When a continuous wave enters an optical fiber with third-order
nonlinearity and the frequency of the incident wave is in the anomalous
group velocity dispersion region of the fiber, 
the incident wave is converted to a pulse train. 
This phenomenon is referred to as (spontaneous) modulation
instability.\cite{agrawal06} 
The modulation instability has been intensively studied 
because it can be used to generate supercontinuum waves, which are useful for
spectroscopy, optical coherence tomography, dense wavelength division
multiplexing, and other applications.\cite{dudley06}  

If modulation instabilities can be generated with high efficiency in other
frequency regions, they could be utilized in 
various other applications. 
This can be achieved by taking advantage of the
energy compression in resonant metamaterials and the scalability of
metamaterial structures. 
Modulation instabilities in 
metamaterials have been studied theoretically in 
three-dimensional magnetoinductive waveguides,\cite{shadrivov06_ph}
negative refractive index media,\cite{wen06} and arrays of silver
nanoparticles\cite{noskov12} and experimentally in composite
right-/left-handed transmission lines.\cite{kozyrev07}

\begin{figure}[tb]
\begin{center}
\includegraphics[scale=1.1]{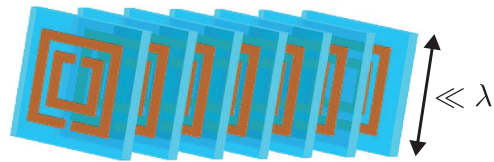}
\caption{(Color online) Schematic of a one-dimensional magnetoinductive waveguide. The
 brown (dark gray) and blue (light gray) 
 parts represent metal and dielectric, respectively. }
\label{fig:MIfig}
\end{center}
\end{figure}

The purpose of this paper is to 
investigate modulation instabilities in a one-dimensional
nonlinear magnetoinductive waveguide numerically and experimentally. 
The magnetoinductive waveguide shown in Fig.\,\ref{fig:MIfig} 
is a subwavelength waveguide that consists
of an array of split-ring resonators.\cite{shamonina08} 
Both the small cross section of the waveguide and the
resonant nature of the split-ring resonators contribute to the
compression of the electromagnetic energy, which leads to 
the enhancement of the nonlinearity. 
Therefore, it is possible to efficiently 
induce modulation instability 
by using nonlinear magnetoinductive waveguides.

\begin{figure*}[tb]
\begin{center}
\includegraphics[scale=1.1]{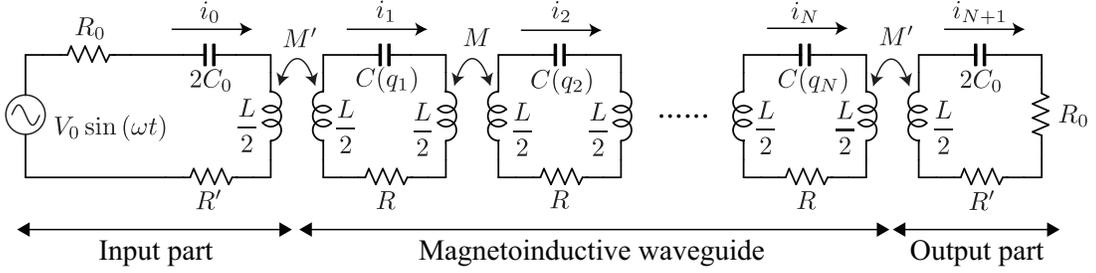}
\caption{Electrical circuit model of a nonlinear magnetoinductive
 waveguide with input and output components. 
Mutual inductances except those between nearest neighbor loops are
 not shown to avoid complexity of the figure. }
\label{fig:MIcircuit}
\end{center}
\end{figure*}

\section{Numerical analysis}

\begin{figure}[tb]
\begin{center}
\includegraphics[scale=1]{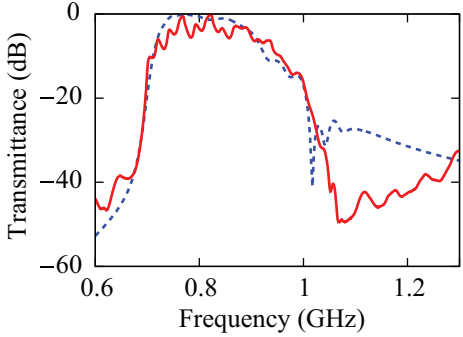}
\caption{(Color online) Linear transmission spectra of the magnetoinductive waveguide
 obtained in the numerical analysis (dashed curve) and experiment (solid
 curve). The transmission spectrum in the numerical analysis 
is normalized to its maximum value. }
\label{fig:transmittance}
\end{center}
\end{figure}

We numerically investigate modulation instability in a one-dimensional 
magnetoinductive waveguide with third-order nonlinearity. 
To simplify the calculation, we analyze an electrical circuit
model of the nonlinear magnetoinductive waveguide, as shown in 
Fig.\,\ref{fig:MIcircuit}. 
The circuit model is composed of a ladder of inductor-capacitor series
resonant circuits that are coupled with each other via mutual
inductances. The capacitors are assumed to exhibit third-order nonlinearity.
Each end of the loop array represents a transmission line
(with a characteristic impedance of $R_0$) and a transducer (with an 
inductance of $L/2$, a
capacitance of $2C_0$, and a resistance of $R^{\prime}$) that couples the
magnetoinductive waveguide to the transmission line. 
The resonant transducers are designed to suppress reflection
at the interfaces between the waveguide and the
transmission lines.\cite{syms10} 
The sinusoidal voltage 
source in the leftmost loop corresponds to an incident
monochromatic wave. 
Applying Kirchhoff's voltage law for the $n$th loop yields the
following equation: 
\begin{equation}
L\frac{\dd i_n}{dt} + R i_n + v_{n} + \sum_{u=1}^U M_u 
\left( \frac{\dd i_{n+u}}{\dd
t} + \frac{\dd i_{n-u}}{\dd t}  \right) =0, 
\label{eq:MIcircuit}
\end{equation}
where $M_u$ is the mutual inductance between the $n$-th loop and the
$(n\pm u)$-th loop and $U$ is the upper limit of $u$ in the calculation. 
The voltages across the capacitors are assumed to be written as
\begin{equation}
v_{n} = \frac{q_n}{C(q_n)} = \frac{q_n}{C_0} + \alpha q_n^3 ,
\end{equation}
where $C_0$ is the linear capacitance, $\alpha$ is a constant that
represents the third-order nonlinearity, and $\dd q_n / \dd t = i_n$. 
The transmission characteristics of the electrical circuit model of the
nonlinear magnetoinductive waveguide can be found by solving
a set of simultaneous differential equation (\ref{eq:MIcircuit}). 
We numerically solve Eq.\,(\ref{eq:MIcircuit}) using 
the fourth-order Runge-Kutta method. 

\begin{figure*}[tb]
\begin{center}
\includegraphics[scale=0.7]{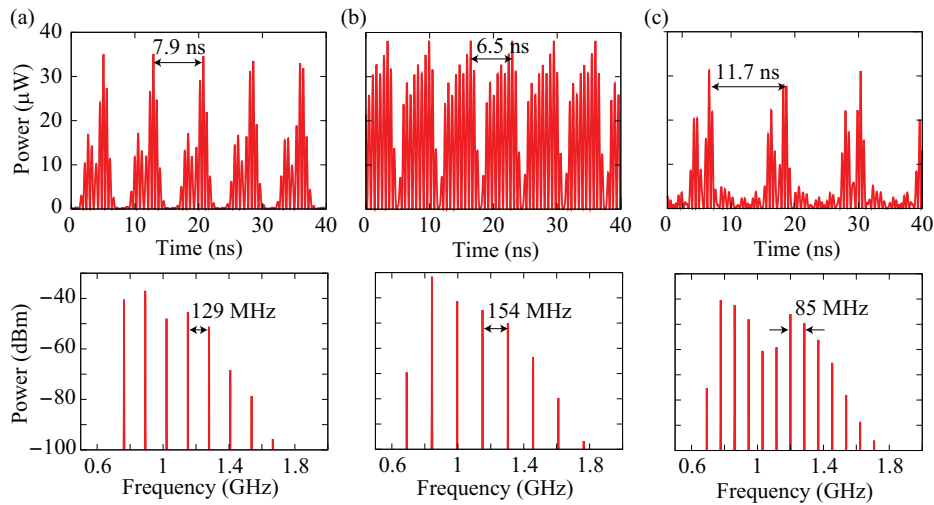}
\caption{(Color online) Numerically calculated waveforms (upper row) and 
spectra (lower row) of the transmitted waves through the electrical
 circuit model of the nonlinear magnetoinductive waveguide for 
(a) $f=1.15\,\U{GHz}$, $P=17.4\,\U{dBm}$, 
(b) $f=1.15\,\U{GHz}$, $P=18.5\,\U{dBm}$, and
(c) $f=1.20\,\U{GHz}$, $P=18.5\,\U{dBm}$. }
\label{fig:sim_data}
\end{center}
\end{figure*}

\begin{figure}[tb]
\begin{center}
\includegraphics[scale=1]{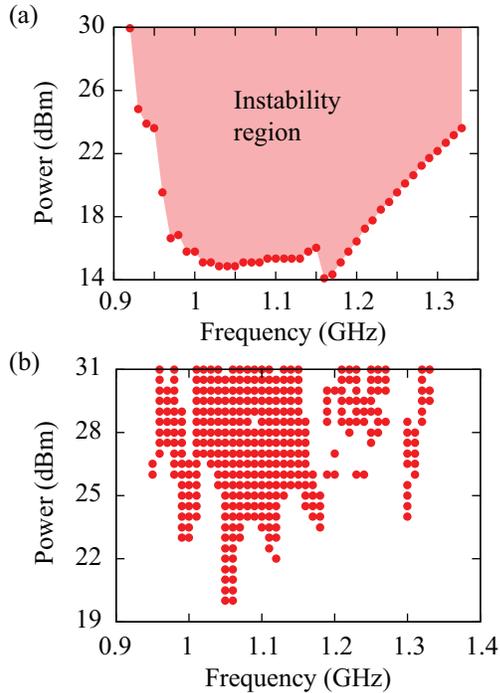}
\caption{(Color online) Region where the modulation instability caused by the instantaneous
 third-order nonlinearity is observed. 
(a) Frequency dependence of threshold value of the source power 
in the numerical analysis. The instability region is shaded in light red
 (light gray). (b) Condition of the incident frequency
 and power in the experiment. }
\label{fig:thre_sim}
\end{center}
\end{figure}

From a preliminary evaluation of the characteristics of 
the nonlinear split-ring resonator used in the experiment (next section), 
we determined the circuit constants as follows. 
The linear capacitance and nonlinear constant were, respectively, set to be 
$C_0=1.08\,\U{pF}$ and $\alpha = 8.64 \,\U{mV/(pC)^3}$ from the datasheet
of the varactor diode (Infineon BBY52-02W) used as the nonlinear capacitor. 
The inductance ($L=27.5\,\U{nH}$) and resistance ($R=7.66\,\U{O}$) were
determined from the linear capacitance, 
measured linear resonant frequency ($920\,\U{MHz}$), and linear 
resonance linewidth. 
The mutual inductances between the two loops
were calculated from Neumann's formula\cite{stratton41}
fitted to the experimental values, which were estimated from the frequency
separation of the two resonance peaks of the two coupled split-ring
resonators.\cite{syms06}    
For simplicity of calculation, 
the mutual inductances between the transducer and the other 
loops were set so
that the coupling coefficient $\kappa^{\prime}_u =
M^{\prime}_u/\sqrt{[L(L/2)]}$ was equal to $\kappa_u = M_u / L$. 
The upper limit of $u$ was set to be $U=10$, 
because $\kappa_{10} < \kappa_1 / 100$ and the $U$ dependence of the
linear transmission spectrum of the electrical circuit model of the
magnetoinductive waveguide was confirmed to be small in the case of $U>10$.
The other parameters were set to be $R^{\prime}=R$, $N=25$,
and $R_0=50.0\,\U{O}$. The dashed curve in 
Fig.\,\ref{fig:transmittance} shows a linear transmission
spectrum derived from the electrical circuit model as the 
power consumption of resistance $R_0$ in the rightmost loop. 
The transmission band in the linear region
is found to be from 0.73\,GHz to 0.88\,GHz. 

Figure \ref{fig:sim_data} shows three examples of 
waveforms and spectra of the
transmitted waves through the electrical circuit model of the nonlinear
magnetoinductive waveguide for different incident frequencies $f$ and
powers $P$. The transmitted waves are
periodically modulated due to the modulation instability. 
While the waveform and spectrum of the transmitted wave are strongly dependent
on the source frequency and power, the separation of the spectral peaks is
of the order of 100\,MHz in every case. 
Note that the generation of nonlinear waves is not derived from
parametric four-wave mixing inside each separated split-ring 
resonator. Only the frequency components that satisfy the phase-matching
condition in the magnetoinductive waveguide can be 
observed in the transmitted waves. That is, the frequencies of the
spectral peaks of the transmitted waves are determined by 
the dispersion relation of the magnetoinductive waveguide. 

Figure \ref{fig:thre_sim}(a) shows the threshold value of the source power for
generating the modulation instability as a function of the source
frequency. The threshold becomes small
in the range of about $1.0$--$1.2\,\U{GHz}$, which is slightly higher
than the linear transmission band. This observation 
is intuitive because, 
as the incident power increases, the resonant frequency of the
split-ring resonator 
shifts to higher frequency due to the third-order 
nonlinearity\cite{poutrina10} 
and the transmission band shifts to higher frequency.

\section{Experiment}

\begin{figure}[tb]
\begin{center}
\includegraphics[scale=0.8]{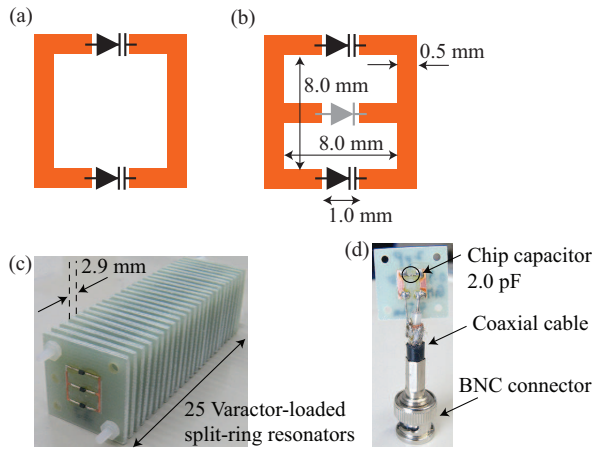}
\caption{(Color online) (a) Varactor-loaded split-ring
 resonator with third-order nonlinearity and (b) that with a faster
 nonlinear response. The split-ring resonator is fabricated using an
 FR-4 printed circuit board, the thicknesses of the copper layer and the
 substrate of which are $35\,\U{um}$ and $1.6\,\U{mm}$, respectively. 
 The black varactor diodes are Infineon BBY-52-02W and
 the gray diode is Rohm 1SS400. Photographs of (c) the
 magnetoinductive waveguide and (d) the resonant transducer connected
 to  a coaxial
 cable. }
\label{fig:srr}
\end{center}
\end{figure}

\begin{figure}[tb]
\begin{center}
\includegraphics[scale=0.85]{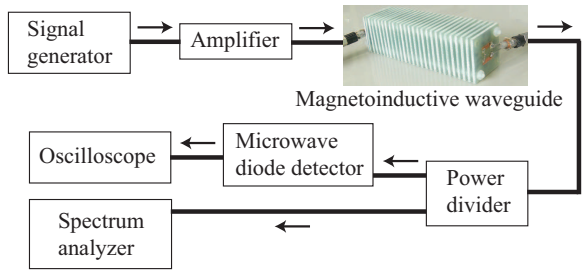}
\caption{(Color online) Experimental setup for observing modulation instability in 
a magnetoinductive waveguide. }
\label{fig:setup}
\end{center}
\end{figure}

In this section we describe the experimental 
observation of modulation instability in the nonlinear
magnetoinductive waveguide in the microwave region. 
Figure \ref{fig:srr}(a) shows a split-ring resonator 
loaded with two back-to-back varactor diodes. The split-ring resonator
exhibits the third-order nonlinearity.\cite{poutrina10} 
Infineon BBY52-02W varactor diodes were used to realize high nonlinearity
with low loss in the split-ring resonator. 
(The quality factor of the split-ring resonator is determined by the
resistance of the varactor diodes rather than the 
radiation loss of the split-ring resonator.)
Since the reverse current of the varactor diode is extremely small, the
nonlinear response of the split-ring resonator is slow due to the
accumulation of minority carriers. To solve this problem, we modified the
structure of the split-ring resonator as shown in
Fig.\,\ref{fig:srr}(b) and introduced another diode (Rohm 1SS400), whose
reverse current is larger than that of the
former diodes. The latter diode has little influence on
the linear characteristics of the split-ring resonator. 
By arranging the 25 split-ring resonators axially with a separation of
2.9\,mm as shown in Fig.\,\ref{fig:srr}(c), 
we constructed a nonlinear magnetoinductive waveguide with
the third-order nonlinearity. 
To couple the magnetoinductive waveguide to coaxial cables, we used 
the resonant transducers\cite{syms10} 
shown in Fig.\,\ref{fig:srr}(d), whose inductance
and capacitance were about half and twice that of 
the varactor-loaded split-ring resonator, respectively.
The solid curve in 
Fig.\,\ref{fig:transmittance} shows the
linear transmission spectrum of the magnetoinductive waveguide
measured using a network analyzer. The frequency range of the 
transmission band agrees well with
that of the numerical calculation. 

The experimental setup for observing modulation instability in the
magnetoinductive waveguide is shown in Fig.\,\ref{fig:setup}. 
A continuous wave was generated from a microwave signal
generator. The wave was amplified and then fed into the
magnetoinductive waveguide. The transmitted wave was split into two halves
by a power divider. One half was detected by a microwave
diode detector to observe the envelope of the
transmitted wave using an oscilloscope. 
The other half was fed into a spectrum analyzer. 

\begin{figure}[tb]
\begin{center}
\includegraphics[scale=0.7]{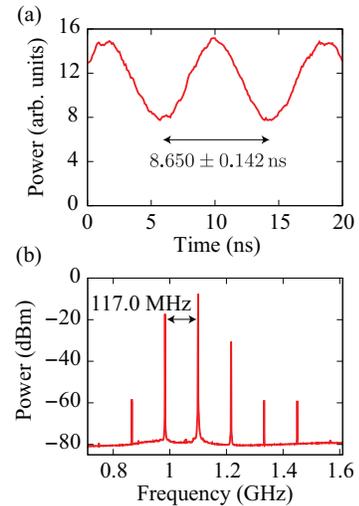}
\caption{(Color online) Characteristics of the transmitted wave for $f=1100\,\U{MHz}$
 and $P=26.5\,\U{dBm}$: (a) waveform of the envelope and (b)
 spectrum. The inverse of the modulation period, $(8.650\,\U{ns})^{-1} =
 115.6\,\U{MHz}$, is consistent with the separation of the spectral
 peaks. }
\label{fig:1100MHz_5.5dBm}
\end{center}
\end{figure}

\begin{figure}[tb]
\begin{center}
\includegraphics[scale=0.7]{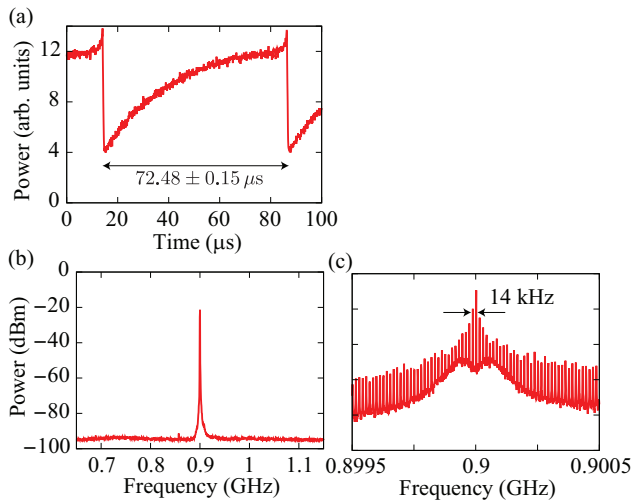}
\caption{(Color online) Characteristics of the transmitted wave for $f=900\,\U{MHz}$
 and $P=26.0\,\U{dBm}$: (a) waveform of the envelope, 
(b) spectrum, and (c) magnified spectrum
 around the peak. The inverse of the modulation period,
 $(72.48\,\U{us})^{-1} = 13.8\,\U{kHz}$, is consistent with the
 separation of the spectral peaks. The difference between the peak values of the
 original and magnified figures is due to the limited resolution of the
 instruments. }
\label{fig:900MHz_5dBm}
\end{center}
\end{figure}

\begin{figure}[tb]
\begin{center}
\includegraphics[scale=0.7]{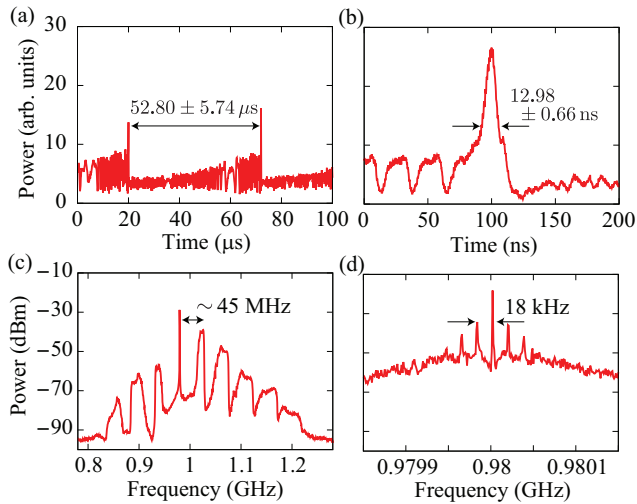}
\caption{(Color online) Characteristics of the transmitted wave for $f=980\,\U{MHz}$
 and $P=26.0\,\U{dBm}$: (a) waveform of the envelope, (b)
 magnified waveform around the peak, (c) spectrum, and (d) magnified
 spectrum
 around the largest peak. The inverse of the modulation period,
 $(52.80\,\U{us})^{-1} = 18.9\,\U{kHz}$, is consistent with the narrower
 separation of the spectral peaks. The difference between the peak values of the
 original and magnified figures is due to the limited resolution of the
 instruments. }
\label{fig:980MHz_5dBm}
\end{center}
\end{figure}

We show three examples of transmitted waves through the nonlinear
magnetoinductive waveguide for different incident frequencies $f$ and 
powers $P$. 
Figure \ref{fig:1100MHz_5.5dBm} gives the measured 
characteristics of the transmitted wave for $f=1100\,\U{MHz}$ and
$P=26.5\,\U{dBm}$. 
In the time domain, 
the transmitted wave is periodically modulated, 
and thus, the modulation instability occurs also in the experiment. 
In the spectral domain, spectral peaks appear 
at spacings of about 100\,MHz, 
which is the same order as in the numerical calculation. 
Figure \ref{fig:900MHz_5dBm} shows the characteristics of the
transmitted wave for $f=900\,\U{MHz}$ and $P=26.0\,\U{dBm}$. 
The modulation period is of the order of $100\,\U{us}$
and the frequency separation between the spectral peaks is of the order
of $10\,\U{kHz}$. These values are of a different order of magnitude than
those obtained from the numerical analysis. Figure \ref{fig:980MHz_5dBm} shows the 
characteristics of the transmitted wave for $f=980\,\U{MHz}$ and
$P=26.0\,\U{dBm}$. Two kinds of spectral
peaks spaced by about $100\,\U{MHz}$ and $10\,\U{kHz}$ are
simultaneously observed. 

\begin{figure}[tb]
\begin{center}
\includegraphics[scale=1]{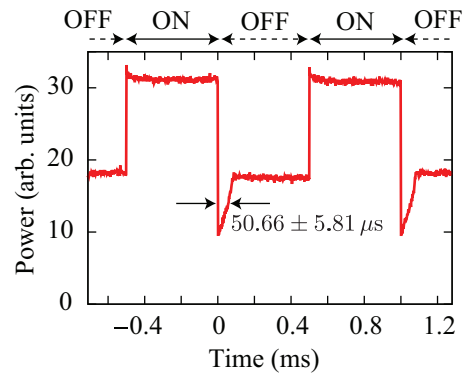}
\caption{(Color online) Measured nonlinear relaxation characteristic of the
 varactor-loaded split-ring resonator. The incident frequency, power of
 the continuous wave, and power of the pulse-modulated wave during the
 pulse-on period are 1100\,MHz, $17.0\,\U{dBm}$, and 31.0\,dBm, respectively. }
\label{fig:relaxation}
\end{center}
\end{figure}

\section{Comparison between numerical analysis and experiment}

We have shown through the numerical calculation and experiment 
that modulation instability occurs in the nonlinear magnetoinductive
waveguide. However, there are some discrepancies between the results of
the numerical 
calculation and experiment. 
In the numerical calculation, there exist only spectral peaks spaced by
about 100\,MHz. In contrast, in the experiment, there exist spectral peaks
spaced by about 10\,kHz as well as 100\,MHz. 
We explore the cause of the difference below. 

The third-order nonlinear response is assumed to occur
instantaneously in the numerical analysis. On the other hand, 
the varactor-loaded split-ring resonators used in the experiment 
show not only the instantaneous nonlinear response but also 
a delayed nonlinear response due to the small reverse current of the
diodes. (Although we made the nonlinear response faster by introducing the
auxiliary diode, the delayed nonlinear response still remains.)
When a third-order nonlinear response of a material occurs
noninstantaneously as well as instantaneously, 
the material will exhibit a Raman gain. 
If we assume that the delayed nonlinear response is described by an
exponential relaxation function, the Raman gain takes a maximum value at
frequencies spaced by the inverse of the relaxation time from the
incident frequency.\cite{stolen89} 
Therefore, we can assume that the
varactor-loaded split-ring resonator exhibits 
a delayed nonlinear response with
a relaxation time that equals the modulation period.

To evaluate the nonlinear relaxation time of the fabricated
split-ring resonator, we measured the envelope of the 
transmitted wave when a superposition of a
pulse-modulated wave (high power) and a continuous wave (low power) 
was injected into a nonlinear
magnetoinductive waveguide composed of one varactor-loaded split-ring
resonator. 
An example of the measured envelope of the transmitted wave is shown in 
Fig.\,\ref{fig:relaxation}. 
During the turn-on transition, the amplitude
of the envelope changes instantaneously. On the other hand, 
we observe a delayed nonlinear response 
with a relaxation time of about $50\,\U{us}$
during the turn-off transition. 
While the relaxation time strongly depends on the incident frequency and
power, the order of the relaxation time agrees with that of the
modulation period for some conditions. 
Therefore, we infer that it is the delayed nonlinear response that
causes the narrowly separated ($\sim 10\,\U{kHz}$) 
spectral peaks---in other words, the 
long modulation period ($\sim 100\,\U{us}$)
in the experiment. 
The asymmetry in the relaxation phenomenon is due to the asymmetry of
the current flow in the diodes. 
If we make a strict electrical circuit model of the split-ring resonator
used in the experiment, we could reproduce the experimentally observed
phenomenon using the numerical calculation shown in the previous
section. However, we do not perform this
simulation here because the strict modeling is beyond 
the scope of the present work. 

From the above discussion, it is found that there exist instantaneous
and noninstantaneous third-order 
nonlinearities in the varactor-loaded split-ring
resonators and each nonlinearity can cause a modulation
instability in the nonlinear magnetoinductive waveguide. 
This implies that 
the modulation instability is caused by 
the instantaneous or noninstantaneous nonlinearity in the case of
Fig.\,\ref{fig:1100MHz_5.5dBm} and Fig.\,\ref{fig:900MHz_5dBm}, respectively, and
by both kinds of
nonlinearities in the case of Fig.\,\ref{fig:980MHz_5dBm}.

In Fig.\,\ref{fig:thre_sim}(b) we show the condition of the incident
frequency and power for generating the modulation instability 
caused by the instantaneous nonlinearity in the experiment
to compare the experimental and numerical results. 
We examined the condition by increasing the incident power from
12.5\,dBm to 
31.0\,dBm with 0.5\,dBm steps in the frequency range 700--1400\,MHz
with steps of 10\,MHz.  
Since the separation of the spectral peaks is of the order of 100\,MHz
and 10\,kHz in the case of the modulation instability caused by the
instantaneous and noninstantaneous nonlinearity, respectively, 
we consider that the modulation instability is caused by the
instantaneous nonlinearity
if the separation of the spectral peaks is larger than 10\,MHz.

While the measured instability condition 
[Fig.\,\ref{fig:thre_sim}(b)] roughly agrees with the
numerical calculation [Fig.\,\ref{fig:thre_sim}(a)], 
there are two minor discrepancies between them.  
First, in the experiment, 
the threshold of the incident power varies rapidly with respect to the
incident frequency. Second,
there exist some conditions where, although the incident power exceeds the
threshold value, modulation
instability does not occur. 
The former difference could be induced because the reflection at the
transducer in the
experiment is larger than that in the numerical calculation. 
The electromagnetic energy density in the
waveguide increases or decreases depending on the incident 
frequency due to the Fabry-Perot effect and thus
the threshold oscillates. 
The latter difference may be caused by the power dependence of the effective 
resistance
of the varactor diode. With increasing incident power, the effective 
resistance increases and
the transmittance in the transmission band decreases (not shown). 
There may be the cases where
the increase of the propagation 
loss becomes larger or smaller than the increase of the gain 
of the modulation instability when the incident power increases. 
The Raman gain in the experiment may also be one of the causes of
the differences.

\section{Conclusion}

We have shown through a numerical analysis of an electrical
circuit model and an experiment in the microwave region that
modulation instability occurs in a magnetoinductive waveguide with
third-order nonlinearity. 
While the condition for
generating the modulation instability in the experiment roughly agrees
with that in the numerical calculation, 
there are some discrepancies
in the characteristics of the modulation instability.
The experimental result implies that 
the difference is due to the noninstantaneous nonlinearity, 
which was not considered in the
numerical analysis. 
This inference could be confirmed by making a strict electrical circuit 
model of the varactor-loaded split-ring resonator
or by a full wave simulation using a
finite-difference time-domain method\cite{taflove05} 
taking
both the instantaneous and noninstantaneous nonlinearities
into account. 
Our study in the microwave region
can be extended to the terahertz and optical regions using
micro- and nanofabrication technologies.

\begin{acknowledgments}

This research was supported by a Grant-in-Aid for Scientific Research on
Innovative Areas (No.\@ 22109004) from the Ministry of
Education, Culture, Sports, Science, and Technology, Japan, and by a
Grant-in-Aid for Scientific Research (C) (No.\@ 22560041) from the Japan
Society for the Promotion of Science. 

\end{acknowledgments}


%


\end{document}